\documentclass[11pt,a4paper]{article}
\usepackage{amsmath,amssymb}
\usepackage{epsfig,graphicx}

\usepackage{graphicx}
\usepackage{wasysym}
\usepackage{epsf}
\usepackage{url}
\usepackage{bm}

\begin{document}

\title{\bf Pulsar timing constraints on narrow-band stochastic signals}
\author{K.~A.~Postnov$^{a b}$\footnote{{\bf e-mail}: pk@sai.msu.ru},
N.~K.~Porayko$^{a b}$\footnote{{\bf e-mail}: porayko.nataliya@gmail.com}
\\
$^a$ \small{\em Sternberg Astronomical Institute} \\
\small{\em Sternberg Astronomical Institute, Universitetsky pr., 13, Moscow 119991, Russia}\\
$^b$ \small{\em Faculty of Physics, Moscow M.V. Lomonosov State University, Leninskie Gory, Moscow 119991, Russia}
}
\date{}
\maketitle

\begin{abstract}
We consider the sensitivity 
of the pulsar timing array (PTA) technique to specific kind of 
narrow-band stochastic signals in nano-Hz frequency range. Specifically, 
we examine the narrow-band signal produced by oscillating gravitational 
scalar potentials in the Galaxy (Gravitational Potential Background), which arise 
if an ultralight massive scalar field is the galactic dark matter. We have performed a
Bayesian analysis of publicly available data on 12 pulsars obtained by the NANOGrav project.
In the monochromatic approximation, the upper limit on the variable 
gravitational potential amplitude is $\Psi_c<1.14 \times 10^{-15}$, 
corresponding to the dimensionless strain amplitude $h_c=2\sqrt{3}\Psi_c < 4\times 10^{-15}$ 
at frequency $f=1.75\times 10^{-8}\mathrm{Hz}$. 
In the narrow-band approximation, the upper limit on the energy density  
of GPB is found to be \(\Omega_{\mathrm{GPB}}<1.27 \times 10^{-9}\) at \(f=6.2\times 10^{-9}\mathrm{Hz}\). These limits are an order of magnitude higher than the theoretically expected values,
if the ultralight scalar field with a mass of $\sim 10^{-23}$~eV is assumed to be 
the galactic dark matter with a local density of $\sim 0.3$~GeV~cm$^{-3}$. 
\end{abstract}
\maketitle

\section{Introduction}
\label{Introduction}
Direct detection of gravitational waves (GWs), predicted by general relativity (GR), 
stands among the principal challenges in experimental astrophysics during the last few decades. 
Measurements of the orbital decay of the binary pulsar J1915+1606  and
similar systems  \cite{2014arXiv1403.7377W}
remain so far the strongest observational evidence for GW emission by space systems.       
Prompt development of GW detectors and projects, including ground-based and space interferometers, pulsar-timing and measurement of the anisotropy of cosmic microwave background, will likely result in the direct detection of GWs in the near future (see recent reviews \cite{lrr-2013-7, lrr-2013-9}). 

A high stability of spin frequency of pulsars, especially old 
recycled millisecond pulsars, allows GW detection in pulsar timing measurements
\cite{Sazsin, Det}.
A pulsar-timing GW detector is represented by two ``free'' masses: 
Earth and a pulsar. In the presence of a GW, 
electromagnetic signal from the 
pulsar, when traveling in perturbed spacetime, 
undergoes the frequency shift. As a result, the variations with particular form in times of arrival (TOA) of pulses from the given pulsar, which are usually measured in pulsar timing experiments, arise \cite{Estabrook}. The main difficulty in detecting, for example, a monochromatic GW using this method is that variations in the pulsar timing residuals can be also caused by inaccurate measurements of the pulsar model parameters and the signal propagation in the interstellar space, therefore the timing residuals should be accurately cleaned from 
many possible contaminating effects. 

The pulsar-timing procedure is sensitive to GWs in the frequency range 
which is limited by the Nyquist frequency (as determined by the duty cycle of the measurements, about two weeks) 
and by the entire time span of the observations (usually several years), i.e. 
\(f_{\text{GW}}\in[10^{-9}\mathrm{Hz};10^{-7}\mathrm{Hz}]\). As the GW detection procedure is determined by properties (amplitude, spectrum, etc.) of the sought signal, the searching strategies usually aim at detecting GWs from  dominant sources in this 
frequency range, for example from supermassive black-hole binaries (SMBHBs) \cite{Jenet}, which are presumably located in galactic centers, 
and from the stochastic gravitational wave background (GWB) produced by the whole population of extragalactic SMBHBs \cite{Sesana} or, likely, by several bright sources above a weak GWB \cite{Bab, BabakSes}. 

The idea of pulsar-timing Arrays (PTAs) was proposed in \cite{1990ApJ...361..300F}. Besides the improvement in sensitivity to deterministic GW signals, the usage of data from several pulsars offers the possibility to search for stochastic GW background by cross-correlating 
timing residuals from different pulsars \cite{Hellings}. Several different PTA projects are currently running: EPTA \cite{Janssen}, PPTA \cite{Manchest}, NANOGrav \cite{Nanoam}, joined in the intrenational project IPTA \cite{IntPuls} (see \cite{2013CQGra..30v4001L} for a review of the PTA techniques).  

In addition to the ``traditional'' GW sources and stochastic backgrounds that can be probed in the PTA frequency range, 
there can be more exotic signals, including, for example, GWs from oscillating string loops \cite{Damour}, GW signals with memory \cite{Postnov, Levinmem}, GWs from massive gravitons \cite{2005PhRvL..94r1102D, Baskaran, Pshirkov, LeeJen}, etc. 

Recently, Khmelnitsky and Rubakov \cite{Khmelitsky} considered a model of an 
ultralight scalar field with bosons mass $m \simeq 10^{-23}-10^{-24} eV$ as a warm dark matter candidate. Ultralight scalar fields as dark matter have been discussed in the literature:
see e.g. Ref. \cite{1985MNRAS.215..575K, 2000PhRvL..85.1158H, 2002PhRvD..65h3514A, 2013arXiv1302.0903S, PhysRevD.62.103517} and references therein. In this model, due to unusually low boson mass, the dark matter density perturbations should be suppressed on subgalactic scales and would behave as the classical cold dark matter on scales larger than the Jeans wavelength $r_{J}=150kpc(10^{-23}eV/m)^\frac{1}{2})$. Moreover, all inhomogeneities are smoothed on scales shorter than the de Broglie wavelength $\lambda = 1/(mv) \simeq 600 pc (10^{-23} eV/m)(10^{-3} c/v)$, where $v$ is the paricle velocity in the Galaxy. As the occupation number of dark matter particles in the galactic halo is huge, the collection of ultra-light particles can be described by a classical scalar field oscillating with frequency $m$. In turn, these oscillations would produce an oscillating pressure at frequency $\omega=2m$ which is averaged to zero on time-scales larger than the period of oscillations. This makes the ultralight scalar field 
effectively non-interacting both with its own particles and particles of the Standard model. However, the pressure oscillations would result in variations in the scalar 
gravitational potentials, which can be probed by the PTA technique 
in a similar way as traditional GWs. 
Note that through a dilatonic coupling 
with the standard model particles, these oscillations could also be probed by atomic clock
experiments \cite{2014arXiv1405.2925A}. 

The effect of the oscillating gravitational potentials on time arrival residulas from 
a pulsar is different from GWs: the response of each detector Earth-pulsar is independent of the source location on the sky, and the scalar field itself is not
an individual source with given angular position. As the distances to pulsars are usually
exceed de Broglie wave of the field ($\sim 600$~pc for the fiducial mass of
the field $m=10^{-23}$~eV), the pulsar signal would propagate through regions 
with uncorrelated field phase, producing a stochastic narrow-band signal in the TOA residuals.
To search for the imprint of the scalar field in pulsar timing data we have used data from NANOGrav Project, which are described in detail in Ref. \cite{Demorest}. 



\section{Pulsar-timing response on the massive scalar field}
\label{signatures}

A scalar field oscillating with frequency $\approx m$ can be represented as
a collection of almost monochromatic ($\Delta \omega/\omega\sim v^2\sim 10^{-6}$) 
plane waves, producing the  
oscillating pressure, and hence, through purely gravitational coupling, the variable 
scalar gravitational potentials $h_{00}=2\Phi$ and $h_{ij}=-2\Psi\delta_{ij}$
(in the Newtonian conformal gauge) at frequency $\omega=2\pi f=2m$. In the weak field approximation two potentials converge to classical Newtonian potential and become equal to each other.
As an electromagnetic signal from a pulsar travels
through the time-dependent spacetime, the irregularity in the
strict periodicity of TOAs of pulses occurs.
Physically, this effect is similar to the classical Sachs-Wolfe effect 
\cite{1967ApJ...147...73S, 2007GReGr..39.1929S}. 

The frequency shift of an electromagnetic wave propagating 
in the variable scalar potential background is \cite{2011iteu.book.....G}:
\begin{equation}
\frac{\nu(t'')-\nu(t')}{\nu(t')}=\Psi(\bm{x_{obs}},t'')-\Psi(\bm{x_{em}},t')-
\int_{t'}^{t''}n_i\partial_i(\Phi+\Psi)dt,
\end{equation}
where $\bm{x_{obs}},t''$ and $\bm{x_{em}},t'$ are the coordinate and time of the receiver and emitter, respectively. The value of the integral term in the above formula is suppressed by the small factor $k/\omega=v\sim 10^{-3}$ 
relative to the amplitude of potentials. Thus, only the variable part of the 
potential  $\Psi_c$, which can be associated with the local dark matter density and the field mass as $\Psi_c\sim \rho_{\text{DM}}/m^2$ \cite{Khmelitsky}, contributes to the pulsar timing signal.

The form of the resulting signal in TOA is:
\begin{equation}
\label{e:R(t)}
R(t)=\frac{\Psi_c}{2 \pi f}\left\{(\sin(2 \pi f t+2 \alpha (\textbf{x}_e))-\sin(2 \pi f (t-D/c)+2 \alpha (\textbf{x}_p))\right\},
\end{equation}
where \(f\) is the frequency, $D$ is the distance to the pulsar, $c$ is the speed of light, \(\alpha(\textbf{x}_e)\) and \(\alpha(\textbf{x}_p)\) are the field phases on Earth and at the pulsar, respectively, and
$\Psi_c$ is the variable potential amplitude to be constrained from the PTA timing analysis. 

The particular feature of this signal is that the signal amplitude does not depend on the angular distance between the source and the pulsar. Below, we will refer to the first and second terms in 
Eq. (\ref{e:R(t)}) as the ``Earth-term'' and ``pulsar-term'', respectively.  

\subsection{Monochromatic approximation}
\label{s:monochrom}

The expected signal is concentrated within 
a very narrow frequency band $\delta f/f\sim v^2\sim 10^{-6}$, which is much 
smaller than the current PTA frequency resolution $\Delta f/f \sim 10^{-4}$, so we can neglect the signal broadening and consider the signal as monochromatic.
In this approximation, the signal is given by Eq. (\ref{e:R(t)}). 

However, the ``pulsar-terms'' add up at different phases due to low accuracy in the distance determination to pulsars. Usually, the ``pulsar-term'' is dropped and treated 
as part of the noise. Here 
we will analyze both cases (including and dropping the pulsar term) in Eq. (\ref{e:R(t)}).
We will denote the effective phase angle due to the pulsar 
\(\theta\equiv \alpha(\textbf{x}_p)-\pi f D/c\), which is individual for each pulsar and is assumed to be uniformly distributed within the interval \([0, 2\pi]\). 

\subsection{Narrow-band approximation}
\label{s:nB}



In this approximation the signal is treated as a narrow-band
stationary stochastic background with power contained within the frequency band $\delta f$ around the central frequency $f$. The narrow-band background can be treated in the same way as stochastic GWB, however,
some differences do arise due to different geometrical structures of
GWs and a variable gravitational potential signal. We begin with comparing the case of the oscillating GPB with classical stochastic GWB \cite{2001PhyU...44R...1G, Baskaran}. 
  
The properties of a stationary statistically homogeneous and isotropic 
GW field can be fully described by the metric power spectrum
$P_h(k)$ per logarithmic interval of wave numbers \(k=2 \pi f/c\):
\begin{equation}
\langle h_s(k^i)h_{s'}^*(k^{'i})\rangle = \delta_{ss'}\delta^3(k^i-k^{'i})
\frac{P_h(k)}{16\pi k^3},
\label{e:h(t)}
\end{equation}
where the angular brackets denote ensemble averaging over all possible 
realizations, the mode functions $h_s(k^i)$ correspond to plane monochromatic waves, and
$s=1,2$ correspond to two linearly independent modes of polarization.

In the case of a
narrow-band signal concentrated within some theoretically prescribed 
interval $\delta k$, 
it is essential to introduce the spectral amplitude $P_0$
\begin{equation}
\label{e:delta}
P_h(k')=\begin{cases}P_0, k<k'<k+\delta k\\
0, \hbox{in other cases}.
\end{cases}
\end{equation}
This allows us to relate $h_c$ and $P_0$:
\begin{equation}
\label{e:P0}
h_c^2= \langle h^2\rangle =P_0\delta f/f\,.
\end{equation}
From above equation we see that the energy parameter is $P_0 \delta f$.

It is also customary to relate the characteristic 
strain amplitude \(h_c(k)\) to the energy density of a stochastic background
per logarithmic frequency interval 
\begin{equation}
\rho_{\mathrm{GWB}}=(16\pi G)^{-1} 4\pi^{2}f^2 h_c^2,
\end{equation}
or, in dimensionless units, 
\begin{equation}
\Omega_{\mathrm{GWB}}=\frac{\rho_{\mathrm{GWB}}}{\rho_{cr}}=\frac{2\pi^2}{3H_0^2}f^2h_c^2=\frac{8\pi^2}{H_0^2}f^2\Psi_c^2,
\end{equation}
where the critical density is 
\(\rho_{cr}=3 H_0^2/(8 \pi G)\) and $H_0$ is the present-day Hubble constant. 


In the PTA data analysis we also need the spectrum $S(f)$ of the TOA residuals produced by 
the sought stochastic signal, which can be obtained via transfer function of the residuals $\tilde R^2(k)$.
For example, in the case of an isotropic GWB for 
the one-sided spectral density of the residuals, we obtain the well-known result
\begin{equation}
S_{\mathrm{GWB}}(f)=\frac{h_c^2}{12\pi^2 f^3}.
\label{SpectrGWB}
\end{equation}
When deriving this formula,
the averaging over the GW tensorial structure and polarization properties 
has been made. Repeating the derivation of the transfer 
function $\tilde R^2(k)$ as in Ref. \cite{Baskaran} for the sought signal from 
oscillating scalar gravitational potential $\Psi_c$ [Eq. (\ref{e:R(t)})], we arrive at 
\begin{equation}
S_{\mathrm{GPB}}(f)=\frac{\Psi_c^2}{\pi^2 f^3},
\label{Spectr}
\end{equation}
which is 12 times as high as Eq. (\ref{SpectrGWB}). 
This independently checks the relation between the equivalent GW characteristic 
strain $h_c$ and the amplitude of the varying potential $\Psi_c$ calculated in 
\cite{Khmelitsky} [see their Eq. (3.9)]: $h_c=2\sqrt{3}\Psi_c$.

As we are working in time domain, 
the knowledge of the covariance function  $C$ of the sought signal is needed.
For a stochastic background, the variance covariance function $C$ 
is related to the signal spectral density \(S(f)\) via the Wiener-Khinchin theorem:
\begin{equation}
C(\tau)=\int^{\infty}_{0}S(f)\cos(\tau f)df\,.
\label{Cint}
\end{equation}
Using the equation for the one-sided spectral density [Eq. (\ref{Spectr})], we get:
\begin{equation}
C(\tau)=\int_0^{\infty} \frac{Q}{f^3} \text{cos} (\tau f) d f=Q(\frac{\tau \text{sin} (f \tau)}{2 f} -\frac{1}{2} \tau^{2} \text{cosIntegral} (f \tau)
-\left.\frac{\text{cos} (f \tau)}{2 f^2})\right|^{f+\frac{\delta f}{2}}_{f-\frac{\delta f}{2}}.
\end{equation}
This can be expanded in Maclaurin series:
\begin{equation}
C(\tau)=\frac{Q}{f^2}\{\text{cos}(f \tau)\left(\frac{\delta f}{f}\right)-\frac{f^2 \tau^2 \text{cos} (f \tau)-12 \text{cos} (f \tau)
-6 f \tau \text{sin}(f \tau)}{24 f^3} \left(\frac{\delta f}{f}\right)^3+O\left(\left(\frac{\delta f }{f}\right)^4\right)\}
\end{equation}
Note that the coefficient before $(\delta f/f)^2$ is zero. 
Due to very narrow frequency range of the GPB signal, 
only the first term is retained. Thus, the covariance matrix $C_{GPB}$ becomes:
\begin{equation}
\label{e:CGPB}
C_{\mathrm{GPB}}(\tau_{ij})=\zeta_{\alpha\beta}\frac{\Psi_c^2 \delta f}{\pi^2 f^3} \mathrm{cos}(f\tau_{ij}),
\end{equation}
where \(\tau_{ij}=2\pi|t_i-t_j|\), \(i\) and \(j\) are indexes of TOA, and \(f\) is the central frequency of the GPB under study. Here \(\zeta_{\alpha\beta}\) is the correlation term between pulsars (\(\alpha\), \(\beta\)). As discussed above, GPB oscillations will induce a sinusoidal signal in 
the TOA of each pulsar with the correlation which takes the simple form
(in contrast, for example, to the case of the GWB from merging SMBHBs):
\begin{equation}
\zeta_{\alpha\beta}=1/2(1+\delta_{\alpha\beta})\,.
\end{equation}
Here the first and second terms arise due to the correlations between 
the pulsar term and the Earth term in Eq. (\ref{e:R(t)}), respectively.

\section{Method of data analysis and data description}
\label{method}

The algorithm of data treatment was implemented in the time domain in Ref. \cite{Levin}, which is commonly applied to unevenly sampled data to avoid the spectral leakage problem.
Generally, pulsar-timing TOA \(\textbf{t}^{arr}\) contains deterministic and stochastic part:
\begin{equation}
\textbf{t}^{arr}=\textbf{t}^{det}(\bm{\beta})+\bm{\delta} \textbf{t}.
\end{equation}
In this equation deterministic part is dependent on the pulsar model parameters \(\bm{\beta}\).

In our analysis the stochastic part \(\bm{\delta} \textbf{t}\) is assumed to include three components. First component is white instrumental noise with the diagonal covariance matrix \(C_{\mathrm{WN}}\). The red intrinsic noise can be characterized by the matrix 
\(C_{\mathrm{RN}}\). The intrinsic pulsar red noise is a challenging problem in the pulsar-timing analysis
because it strongly affects the PTA sensitivity
to GW signals. The nature of this type of noise is not completely clear and can be 
related, for example, to irregular momentum exchange between the superfluid component and the 
neutron star crust, or to fluctuations of the electron density in the interstellar medium \cite{Lommen}. The red-noise spectrum is usually assumed to have a power-law form
\begin{equation}
S(f)=\frac{A^2}{12 \pi^2} f_0^3 \left(\frac{f}{f_{0}}\right)^{-\gamma},
\end{equation}
which is dependent on two parameters \(A_{RN}\) and \(\gamma_{RN}\) individual for each pulsar. 

The last part corresponds to the oscillating GPB  
\(C_{\mathrm{GPB}}\) (in the narrow-band approximation). 
Thus, the covariance matrix of the pulsar-timing data 
 \(\bm{\delta t}\) can be expressed as \(C=C_{\mathrm{WN}}+C_{\mathrm{RN}}+C_{\mathrm{GPB}}\).
The specific form of the white and red noise covariance matrices are discussed in \cite{LevinH}. 

In the time domain, we have implied the likelihood function technique to estimate the parameters of both the deterministic part and colored pulsar noise spectra \cite{Levin}. After marginalization over unwanted pulsar model parameters $ \bm{\beta}$, in the Bayesian approach we obtain the following expression for likelihood function, 
assuming the Gaussian distribution of \(\bm{\delta t}\):
\begin{equation}
\begin{split}
P(\boldsymbol{\delta} \textbf{t}|\bm{\phi})=\frac{1}{\sqrt{(2\pi)^{(n-m)}\mathrm{det} (G^{\mathrm{T}} C G)}}\\
 \mathrm{exp}(-\frac{1}{2}\bm{\delta} \textbf{t}^{\mathrm{T}} G (G^{\mathrm{T}} C G)^{-1} G^{\mathrm{T}} \bm{\delta} \textbf{t}).
\end{split}
\end{equation}
Here \(n\) is the dimension of \(\bm{\delta} \textbf{t}\), $m$ is a whole number of the unwanted parameters, \(\bm{\phi}\) is the noise parameter vector, and \(G\) refers to the product of the so-called ``design matrix'' that can be obtained using the design matrix plugin of the {\small TEMPO2} software \cite{LevinH, Hobbs}. 

In the case of narrow-band stochastic process the spectral density (thus, the covariance matrix) of the sought gravitational potential is parameterized by Eq. (\ref{Spectr}) and depends on the dimensionless power of the stochastic background $ \Omega_{\mathrm{GPB}}$ and central frequency $f$.

While searching for the deterministic signals (\(\ref{e:R(t)}\)), we have used the logarithmic likelihood ratio function which refers to the probability whether the signal is present or absent. In this case the function to be optimized depends on the amplitude \(\Psi_c\) and the Earth phase \(\alpha(\textbf{x}_{e})\) of the scalar field when using the Earth term only, and on \(N+2\) parameters: the amplitude \(\Psi_c\), the Earth phase \(\alpha(\textbf{x}_{e})\) and  the phase \(\theta_\beta=\alpha(\textbf{x}_{p}^{\beta})-\pi f D_{\beta}/c\), if both the Earth and pulsar terms are retained in the analysis. 

The procedure of setting an upper limit on \(\Psi_c\) (\(\Omega_{\mathrm{GPB}}\)) as a function of the 
central frequency $f$ is similar in both cases. We split the entire interesting frequency range into small bins per logarithmic scale [\(\delta f/f\simeq 0.03 \ll 1/(5\,\mathrm{yrs})\)]. In each bin we construct a long enough chain using Markov-chain Monte-Carlo method \cite{Newman} in order to restore the posterior distribution of \(\Psi_c\) (\(\Omega_{\mathrm{GPB}}\)). The prior distribution for red-noise parameters $A$ and $\gamma$ was assumed to be multivariate normal, while for both amplitudes $\Psi_c$ and $\Omega_{\mathrm{GPB}}$ the prior distributions were chosen to be uniform.

The obtained posterior distribution for $\Psi_c$ and $\Omega_{\mathrm{GPB}}$, which turned out to be close to uniform one, was used to set upper limit on the amplitude. In other words, we estimate the Bayesian posterior distribution of the amplitude with MCMC method and assume that 
the amplitude of the probable signal with 95 \% probability is limited by 0.95 quantile of the posterior distribution \cite{2011MNRAS.414.3117V}, taking into account uniform prior distributions.
Results of this analysis are presented below in Sec. \ref{results}.

Methods described above were applied to real data from the NANOGrav Project. The observations were conducted during a 5-yr period from 2005 to 2010 using two radio telescopes, the Arecibo Observatory and NRAO Green Bank Telescope. The data are described  in detail in Ref. \cite{DemorestNanoG} and are publicly available at \footnote{\url{http://www.cv.nrao.edu/~pdemores/nanograv_data/}}. To increase the signal-to-noise ratio in each observation, we have compressed the data by ``daily averaging'' TOAs \cite{Lommen}. Data from 12 pulsars have been analyized. Four of them (J1713+0747, J2145-0750, B1855+09 and J1744-1134) show a weak red-noise component that have been estimated using the MCMC method. The post-fit residuals were obtained with the {\small TEMPO2} software \cite{Hobbs}. The additive and multiplicative factors (EFAC, EQUAD) were not been added to the ``free parameter'' template.

\section{Results and Discussion}
\label{results}

The nature of dark matter is unkown. One of the possible dark matter candidate in our Galaxy can be an ultralight scalar field (e.g. \cite{Khmelitsky}). In this model, the dimensionless 
amplitude of the variable gravitational potential produced by
the oscillating massive scalar field \(\Psi_c\) is related to the 
local galactic dark matter density $\rho_{\text{DM}}$ and the field mass $m$ as:
\begin{equation}
\label{e:modelpsi}
\Psi_c=\pi  \frac{G\rho_{\text{DM}}}{(\pi f)^2}\approx 
10^{-16}\left(\frac{f}{10^{-8}\text{Hz}}\right)^{-2}
\approx 4.3\times 10^{-16}\left(\frac{m}{10^{-23}\text{eV}}\right)^{-2}
\left(\frac{\rho_{DM}}{0.3\text{GeV}\,\hbox{cm}^{-3}}\right)\,.
\end{equation} 
This field cannot be revealed using common telescopes aimed at the detection of electromagnetic signals, but can be probed with the pulsar-timing tool. 


Working in the time domain we have applied the Bayesian approach developed in Ref. \cite{Levin} 
to search for imprints of the oscillating gravitational potential in pulsar-timing data from the NANOGrav project. In our analysis we took into account the red noise for 4 pulsars in the array. The red-noise parameters were estimated using the MCMC method.  In the data analysis, we 
examined three possible signal types: 
a monochromatic deterministic GPB with the Earth term only, a monochromatic GPB 
including both the Earth and pulsar terms, and a narrow-band stochastic GPB. In all cases we 
obtained upper limits on the signal amplitude \(\Psi_c\) (or \(\Omega_{\mathrm{GPB}}\)) as a function of frequency $f$. The best sensitivity is reached in the case of the monochromatic signal using both the Earth and pulsar terms.

\begin{figure}[htp]
\centering
\includegraphics[scale=1.20]{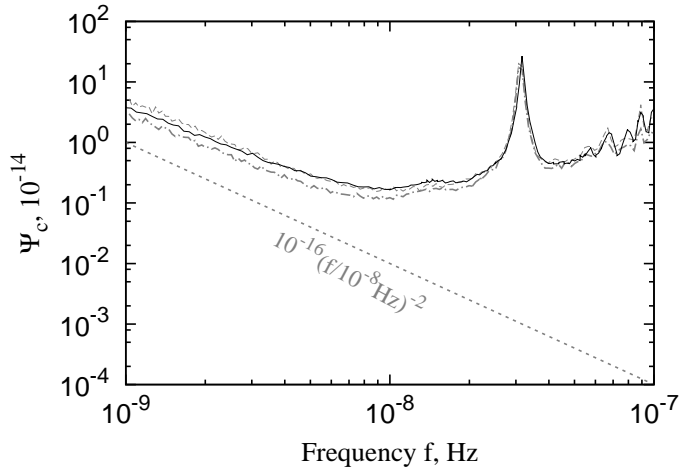}
\caption{An upper limit on the amplitude of the variable gravitational
potential \(\Psi_{c}\) due to the massive scalar field oscillations as a function of the central frequency $f$. Shown is the case of the narrow-band signal approximation (the black line),  the monochromatic signal approximation with the Earth term only (the thin gray dashed line), and
the monochromatic approximation 
using both the Earth and pulsar terms (the gray dashed-dot line); the lines are shown for the 95\% confidence level. Data from eight pulsars from the NANOGrav Project with white-noise rms residuals were used.
The dashed line shows the model amplitude (\ref{e:modelpsi}).
}
\label{result8}
\end{figure}

\begin{figure}[htp]
\includegraphics[width=.49\textwidth]{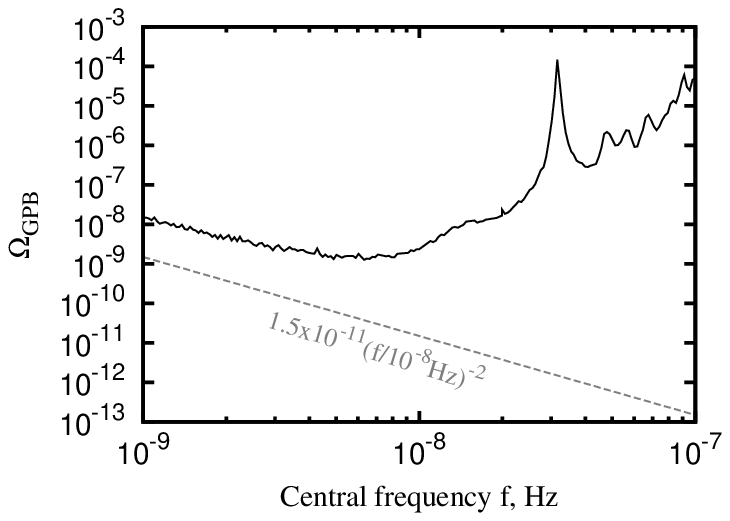}
\hfill
\includegraphics[width=.49\linewidth]{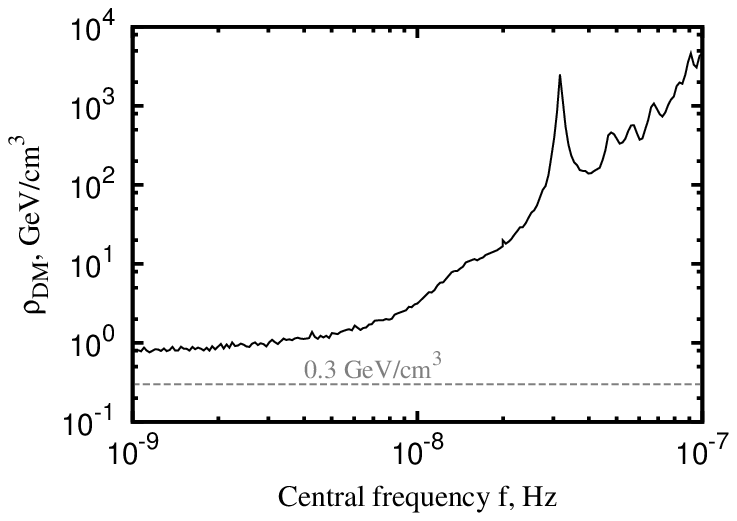}
\caption{
Left panel: An upper limit on the \(\Omega_{\mathrm{GPB}}\) 
of the ultralight scalar field as a function of frequency $f$; the solid curve 
corresponds to the 95\% confidence level. The dashed line shows the model value. 
Right panel: The same 
limit in terms of the local dark matter density \(\rho_{\text{DM}}\). 
The dashed line shows the local galactic dark matter density 0.3~GeV~cm$^{-3}$. 
Data from 12 pulsars from the NANOGrav Project have been used. 
}
\label{result12}
\end{figure}
Our upper limits are based on the Bayesian estimation of the signal amplitude.
The Bayesian estimates can be biased because they do depend on the data realization. 
To obtain non-biased estimates (close to the true parameter values), 
it is necessary to make averaging over data realizations \cite{LevinBook}
\begin{equation}
\hat{\theta}=\int_{X_n} \hat{\theta}_{\textrm{Bayes}} (\textbf{x}) W(\textbf{x}) d \textbf{x},
\end{equation}
where $W(\textbf{x})$ is the ``model evidence''.
In this case the estimates will be independent on the concrete realization of the data. We postpone the investigation of this effect for future work.

In the narrow-band approximation, the stringent limit on the field amplitude  
is \(\Psi_c<1.14 \times 10^{-15}\), which corresponds to the equivalent characteristic
dimensionless strain $h_c=2\sqrt{3}\Psi_c< 4\times 10^{-15}$ at \(f=1.75\times10^{-8}\mathrm{Hz}\) (see Fig. \(\ref{result8}\)). In this approximation 
the power spectral density of the GBP was assumed to have a delta-like form (\ref{e:delta}). Using a flat prior in  the logarithmic scale, we numerically estimated the posterior distribution of the signal power in each frequency bin to set an upper limit on the GPB (in terms of $\Omega_{\mathrm{GPB}}$). In this case, the stringent upper limit is \(\Omega_{\mathrm{GPB}}<1.27 \times 10^{-9}\) at \(f=6.2\times 10^{-9}\)~Hz, which corresponds
to $\Psi_c< 1.5\times 10^{-15}$ (see Fig. \(\ref{result12}\)). The sensitivity curves are similar in both monochromatic and narrow-band cases due to a particularly narrow frequency range of the stochastic signal [less than one frequency bin $\Delta f\sim 1/(5\, \hbox{yr})$]. However, the narrow-band approach can be generally used to search for narrow-band stochastic signals of different origin.

The obtained limits are an order of magnitude higher than the theoretically expected values
if the ultralight scalar field with a mass of $\sim 10^{-23}$~eV is assumed to be 
the galactic dark matter with a local density of $\sim 0.3$~GeV~cm$^{-3}$. 

\section*{Aknowledgements}
The authors thank S. Babak, M. Pshirkov, V. Rubakov and A. Gusev for discussions 
for useful notes. 
The use of the publically available NANOGrav PTA data is acknowledged.  
The work is supported by the Russian Science Foundation grant 14-12-00203.

\bibliographystyle{unsrtmod}
\bibliography{gw.bib}

\end{document}